# Interfacial two-dimensional oxide enhances photocatalytic activity of graphene/titania via electronic structure modification

**Dario De Angelis**[a,†], **Francesco Presel**[a,†,h], **Naila Jabeen**[a,b,c], **Luca Bignardi**[d,i], **Daniel Lizzit**[d], **Paolo Lacovig**[d], **Silvano Lizzit**[d], **Tiziano Montini**[e], **Paolo Fornasiero**[e], **Dario Alfè**[f], **Alessandro Baraldi**[a,d,g],*

[a] Department of Physics, University of Trieste, via Valerio 2, 34127 Trieste, Italy.

[b] Nanosciences & Catalysis Division, National Centre for Physics, Islamabad 44000, Pakistan.

[c] International Centre for Theoretical Physics, Strada Costiera 11, 34151 Trieste, Italy.

[d] Elettra-Sincrotrone Trieste, S. S. 14, km 163.5 in AREA Science Park, 34149 Trieste, Italy.

[e] Department of Chemistry and Pharmaceutics, University of Trieste, and INSTM, via L. Giorgieri 1, 34127 Trieste, Italy.

[f] Department of Earth Sciences, Department of Physics and Astronomy, Thomas Young Centre@UCL, London Centre for Nanotechnology, University College London, Gower Street, WC1E 6BT London, United Kingdom.

[g] IOM-CNR, Laboratorio TASC, S. S. 14 km 163.5 in AREA Science Park, 34149 Trieste, Italy.

[†] These authors contributed equally.

* Corresponding author. E-mail: alessandro.baraldi@elettra.eu (Alessandro Baraldi)

[h] Present address: Department of Physics, Technical University of Denmark, 2800 Kgs. Lyngby, Denmark.

[i] Present address: Department of Physics, University of Trieste, via Valerio 2, 34127 Trieste, Italy.

**Abstract**

A two-dimensional layer of oxide reveals itself as a essential element to drive the photocatalytic activity in a nanostructured hybrid material, which combines high-quality epitaxial graphene and titanium dioxide nanoparticles. In particular, it has been revealed that the addition of a 2D Ti oxide layer sandwiched between graphene and metal induces a p-doping of graphene and a consistent shift in the Ti *d* states. These modifications induced by the interfacial oxide layer induce a reduction of the probability of charge carrier recombination and enhance the photocatalytic activity of the heterostructure. This is indicative of a capital role played by thin oxide films in fine-tuning the properties of heterostructures based on graphene and pave the way to new combinations of graphene/oxides for photocatalysis-oriented applications.

# 1. Introduction

Carbonaceous materials have been widely employed as fundamental building blocks for heterostructures with catalysis- and energy-oriented applications [1-5]. Graphene in particular, owing to its unique electronic and transport properties, is in the spotlight to improve the efficiency of photoabsorbers in photocatalysis and photo-electrocatalysis, which are fundamental building blocks towards a sustainable economy [6]. Since the pioneering work of Williams et al. [7], who developed the first titania-graphene (Gr) nanocomposites, large efforts have been devoted to employ this 2D material, also in its 3D form [8], in combination with oxides ($WO_3$ [9,10], ZnO [11]) or with other semiconductors ($C_3N_4$ [12], CdS [13]) in order to reduce the electron-hole (e-h) recombination rate, by exploiting its high charge-carrier mobility; in order to modify the band alignment, to reduce overpotentials, and in order to reduce the band gap, thus allowing electrons to be excited into the conduction band by visible light [14]. The latter process can occur following two pathways, either an electron excitation from the valence to the conduction band of Gr followed by transfer of the electron into the semiconductor unoccupied states, or a direct excitation from occupied Gr states to the absorbers conduction band. Indeed, the combination of oxides and Gr has been demonstrated experimentally to improve the degradation of organic molecules [15], extend the light absorption range and enhance the charge separation properties [16 – 18], not only when using oxides in form of thin films or nanoparticles (NPs), but also in other forms such as nanowires [19] and nanotubes [20].

It is important to highlight that most of these results have been obtained using Gr-oxide (GO) or reduced-Gr-oxide (RGO) [21], which are characterized by poorer transport properties than high quality Gr, due to the presence of structural defects and functional groups, which are also known to be centers of e-h recombination. This issue strongly limits the photocatalytic enhancement achieved when coupling RGO with semiconductors, to the extent that it has been suggested that GO cannot provide truly new insight into the fabrication of high-performance photocatalysts, in particular when its effect is compared to other carbon-based materials such as carbon nanotubes, fullerenes and activated carbon [22], and points out the importance of using Gr of high structural quality instead.

In this respect, chemical vapor deposition on metal surfaces has proven to be effective for the synthesis of large Gr flakes with a low density of vacancies, domain boundaries and impurities [23]. Moreover, depending on the substrate of choice, this method allows to tune the interaction strength between Gr and the metal substrate [24]. In particular, the supporting substrate plays a fundamental role on the degree of Gr doping and therefore represents an opportunity to tune the properties of Gr-titania hybrid materials by modifying its structure and composition. A powerful strategy to achieve this goal is the intercalation at the Gr-metal interface of any light atoms [25 – 27], molecules [28, 29], alkali- [30], noble- [31] or transition-metals [32] and oxides [33, 34].

In this work we discuss the growth, the structural/electronic characterization and the photocatalytic activity trends of a novel nano-architecture, which has been designed with the aim to investigate the role of the substrate below Gr on the electronic band structure and alignment of the supported photoabsorbers. By using Gr grown on Ir(111), as a prototype of quasi free-standing system, we have prepared a novel interface consisting of TiO$_2$ NPs supported by a layered structure based on high-quality epitaxial monolayer Gr, two-dimensional titanium-oxide and metal.

In order to disentangle the role of Gr and its doping level, we synthesized and characterized several nanoarchitectures, where titania was supported on differently doped Gr, as well as a control system without Gr supporting the titania. In particular, we prepared the three systems shown in the bottom part of Figure 1, corresponding to TiO$_2$ supported on (i) a bare Ir(111) surface (TM, Figure 1d), (ii) Ir-supported Gr (TGM, Figure 1e) and (iii) titanium-oxide-supported Gr (TGTM, Figure 1f). Lastly, the effects of tuning the electronic structure of TiO$_2$ nanoparticles on its catalytic activity have been investigated by using our system as a model catalyst for the hydrogen evolution reaction (HER). Our experiment shows that the doping of Gr greatly affects the HER reaction rate.

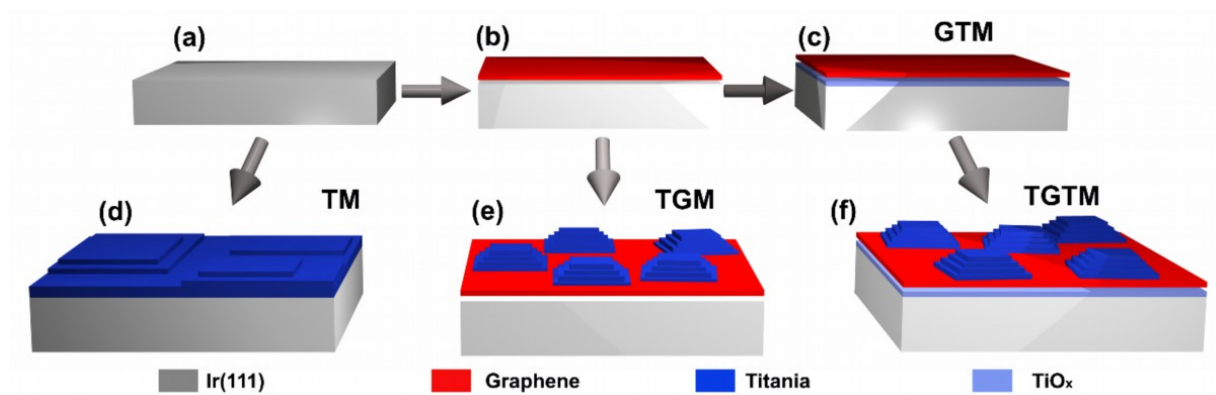

**Figure 1.** Schematic illustration of the systems prepared in the present work. a) Clean Ir(111) surface (M); b) Graphene grown on Ir(111) by means of C$_2$H$_4$ chemical vapor deposition (GM); c) Graphene on a TiO$_x$ interface layer obtained by Ti intercalation and oxidation (GTM); d) TiO$_2$ grown on Ir(111) (TM); e) TiO$_2$ nanoparticles grown on Gr/Ir(111) (TGM) and f) TiO$_2$ nanoparticles grown on Gr/TiO$_x$/Ir(111) (TGTM).

The possibility to adopt similar layered nano-architectures by using the large number of available ultra-thin surface-oxides [35, 36] or by supporting Gr on other two-dimensional materials opens new pathways for the improvement of energy-based applications of Gr [37] and for the design of a novel generation of photocatalytic materials, both for individual use or in tandem cells.

## 2. Results and discussion

The preparation stages of the nanostructures are depicted in Figure 1. In particular, the Gr/Ir interface (Figure 1b) was obtained by growing monolayer Gr on the clean metal surface (Figure 1a) through a well-established synthesis method [38]. It is important to highlight that while the growth parameters (such as substrate temperature and precursor gas pressure) are specific to the Ir substrate chosen for our model system, high quality Gr can be likewise be grown on a wide variety of metal surfaces, including common materials such as Cu and Ni [24]. Eventually, the Gr/oxide interface (GTM, Figure 1c) was obtained in two steps by intercalation of 0.5 monolayers (ML) of Ti at the Gr/metal interface, followed by oxidation [29,32], as described in the Methods section. Again, this procedure is not specific to the substrate used, but can be likewise applied to a wide variety of Gr/metal interfaces [32].

The morphology of $TiO_2$ grown on Gr/$TiO_x$/Ir was characterized by STM (Figure 2a and b), which shows that the surface is homogeneously covered by round-shaped NPs, separated by regions of bare Gr. The average surface corrugation is about 10 Å, as illustrated by the scan line profile reported in Figure 2b, and it is comparable to the value found for $TiO_2$ NPs grown on Gr/Ru(0001) [39]. The size distribution of the particles (Figure 2c), which ranges from 3 to 30 nm, is peaked at 9 nm, comparable to the value found for metallic Ti NPs grown on Gr [40]. A similar distribution of the NPs' sizes was also found for the TGM system, indicating that the presence of the two-dimensional Ti-oxide interfacial layer does not affect the size and distribution of the Gr-supported NPs. In order to have a reference sample to assess the role of graphene in the heterostructure, we have also deposited $TiO_2$ directly on Ir(111) (TM system, figure 1d) with the same method used for the NPs growth on TGM and TGTM structures. After Ti was deposited and oxidized on clean Ir(111), we could observe the presence of a $TiO_2$ layer.

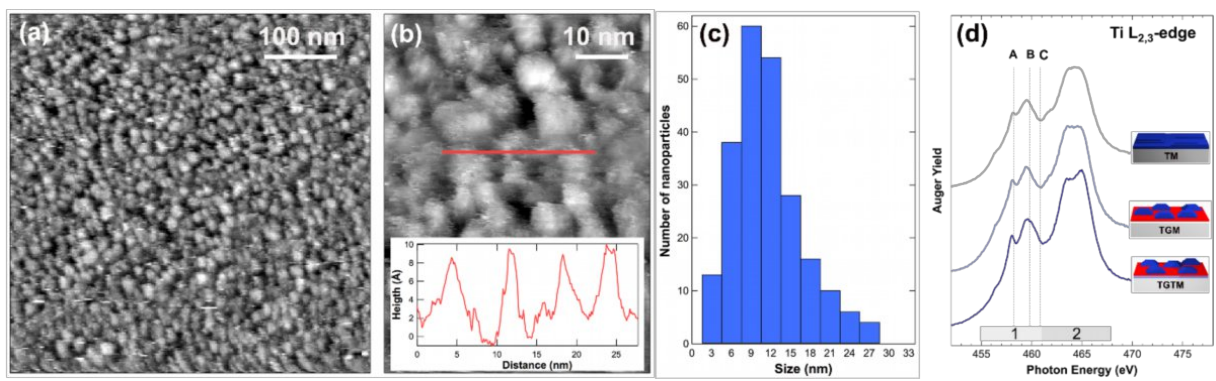

**Figure 2.** Structural characterization of selected systems. a) STM images of TGTM, 500×500 nm$^2$; b) STM images of TGTM, 100×50 nm$^2$, bias = 2.1 V, current = 0.42 nA). The inset shows the height profile measured along the red line; c) Size distribution of NPs obtained from a); d) NEXAFS spectra measured at the Ti L$_3$ and L$_2$ absorption edges for the TM, TGM and TGTM nano-architectures. The energies at which the most prominent features A, B and C are usually found in the spectra of bulk titania samples are indicated with dotted lines.

The local crystal structure of the Ti-oxide particles was investigated by analyzing NEXAFS spectra at the Ti L$_3$ (455 – 461 eV-region 1) and L$_2$ (461 – 468 eV-region 2) thresholds (Figure 2d). Both regions show an intense and structured pre-edge feature. The fine structure is due to the splitting of the Ti $d^*$ bands due to the crystal symmetry. For this reason, the energy value of the maxima is a fingerprint of the various existing TiO$_2$ crystal structures, thus allowing for example to distinguish the rutile and the anatase phases in titania [41]. More specifically, in our case the L$_3$ maxima are found at 458.0 eV (A) and 459.8 eV (B), which indicates that the local crystal structure in our TiO$_2$ NPs is anatase. Interestingly, TiO$_2$ NPs are commonly of the anatase form [42,43], which is generally considered a better photocatalyst than rutile, the most stable polymorph of TiO$_2$, because of its longer carrier lifetime, longer e-h diffusion length, higher surface area and higher carrier mobility.

A very important characteristic of TiO$_2$ NPs is their oxidation state, which is related to the density of O vacancies, which have been proven to play an essential role in determining the photocatalytic activity of titania. In order to quantify their density, we have identified the oxidation states of titanium in each nanoarchitecture by a quantitative analysis of the HR-XPS spectra of the Ti *2p* levels (shown in Figure 3).

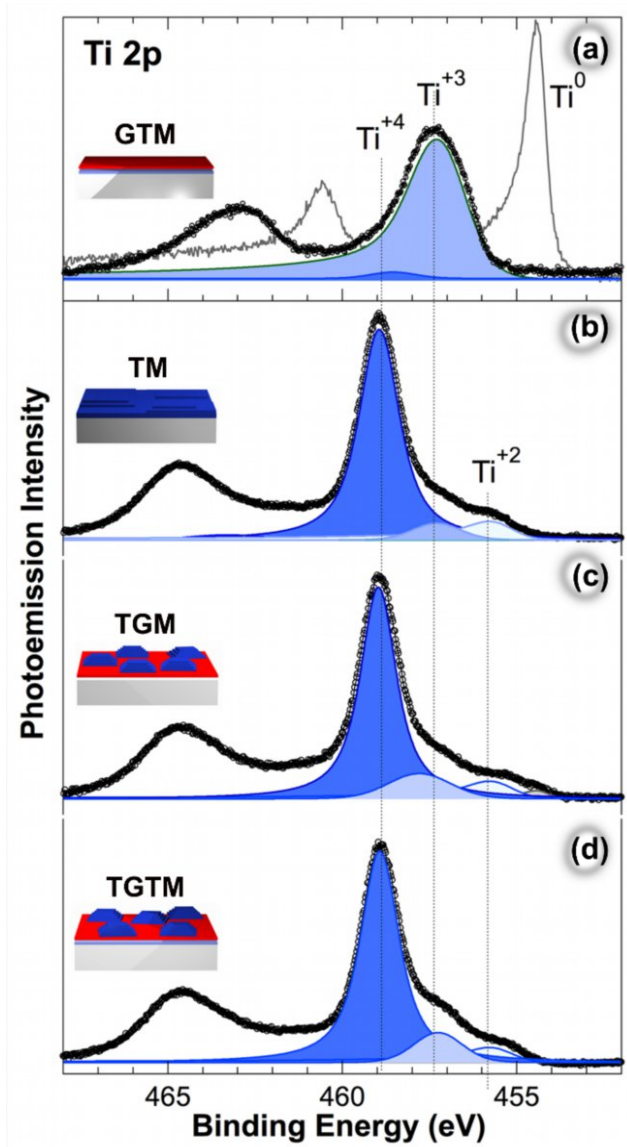

**Figure 3.** Ti *2p* photoemission spectra. $2p_{3/2}$ spectra together with deconvoluted spectral components are reported for selected nano-architectures (hν = 550 eV). The spectrum of the as-deposited metallic titanium, before oxidation, is shown in the top graph for comparison (grey curve).

The Ti *2p* spectrum corresponding to the titanium intercalated layer is shown in Figure 3a. The $2p_{3/2}$ spectral region is characterized by a main peak a binding energy (BE) of about 457 eV and an additional feature at about 459 eV, which can be associated to $Ti^{3+}$ and $Ti^{4+}$ oxidation states, respectively [44]. The Ti *2p* spectrum corresponding to the intercalated metallic Ti layer before oxidation, characterized by the $Ti^0$ component at 454.35 eV, is also shown for comparison. The component with the highest spectral weight in GTM is the one originated by $Ti^{3+}$. The quantitative analysis of the Ti *2p* core level indicates that oxygen to Ti atoms are in a ratio of about 2:3 for GTM. Interestingly, this corresponds to the $TiO_x$ layer (x< 2) that was observed for an incommensurate oxidized titanium film with hexagonal geometry grown on Pt(111) at 670 K [45], i.e. the same temperature we have employed in the oxidation step. The presence of a small amount of Ti atoms in the higher +4 oxidation state might be due to local defects such as additional O atoms at the interface between this layer and iridium, or a local bilayer structure with the stoichiometry of titania.

Following this spectroscopic characterization of the buried Ti-oxide/Ir interface, we evaluated the oxidation state of the three surfaces obtained after the $TiO_2$ growth, whose Ti *2p* spectra are shown in Figure 3(b, c, d). For the case of the TGTM, also the components originating from the buried interface below Gr were included in the fit. In each of the three systems, the main peak at about 459 eV is attributed to Ti atoms in +4 oxidation state. Additional, low-intensity components are observed at lower BE and are a fingerprint of the presence of oxygen vacancies. In each of the three systems, the relative spectral weight of the $Ti^{4+}$ species is between 75 and 82%.

This outcome was combined with the results from resonant photoemission spectroscopy (RESPES) experiments (see Supporting Information) in order to directly compare the density of oxygen defects among the TM, TGM and TGTM nanoarchitectures. In resonant conditions, a spectral feature associated to the presence of oxygen vacancies is easily measurable at about 1.1 eV BE [46]. Its intensity difference between TGM and TGTM is less than 1%, showing that the particles grown on the two substrates are indistinguishable from this point of view. For the TM interface, the photoemission signal is generally higher, but the relative intensity of the 1 eV feature does not show significant differences, implying that the density of O vacancies does not significantly differ in the three different systems. RESPES measurements were also used to extract the value of the band gap of titania, using the method proposed by Das et al. [47] The value of 4.3 eV, which is overestimated when compared to the one obtained using other methods, mainly due to the errors introduced by Koopmans' approximation and to final state effects, is in good agreement with the one we measured for bulk anatase using the same method. This suggests that the size of the NPs is large enough to consider their band-gap the same as bulk-like anatase titania.

HR-XPS experiments returned important information about the charge transfer occurring between Gr and both its substrate and the NPs, whose composition influences the doping level of Gr and affects the BE value of the C $1s$ core level peak [48]. Figure 4a shows the C $1s$ core level of Gr/Ir and Gr/TiO$_x$/Ir before and after the growth of TiO$_2$ NPs. In particular, the C $1s$ spectrum of Gr/Ir is centered at 284.12 eV, indicating a slight p-doping [49]. The growth of titania NPs on Gr/Ir (TGM spectrum) results in a shift towards higher BE. The growth of an oxide layer on top of Gr is indeed expected to result in an increase of the C $1s$ BE because of a shift in the Fermi level, as reported for the case of yttria-covered epitaxial Gr grown on Pt(111), which results in Gr n-doping [50]. After intercalation of a 0.5 ML of Ti below Gr, we found the same type of doping, but even to higher extent, thus resulting in a C $1s$ component shifted to higher BE (284.7 eV), which reminds the case of Ti deposited on SiC-supported Gr [51]. A corrugation induced in Gr in strongly interacting systems by the lattice mismatch between Gr and the metal substrate accounts for the double C $1s$ component observed [32,52,53]. However, after oxidation of the intercalated Ti layer, the C $1s$ peak shifts to an even lower binding energy with respect to Gr/Ir (about -0.50 eV), indicating that, in this configuration, Gr is strongly p-doping. The presence of interfacial oxide layers, such as alumina [33], in between Gr and the supporting metal substrate, also resulted in a considerable p-doping even though in these cases the core level shifts were smaller. Last, the deposition of titania on top of Gr/TiO$_x$/Ir nanoarchitecture (TGTM spectrum) shifts again the C $1s$ BE towards higher values, only partially compensating the effect of p-doping due to the presence of the interfacial oxide (C $1s$ BE = 283.95 eV).

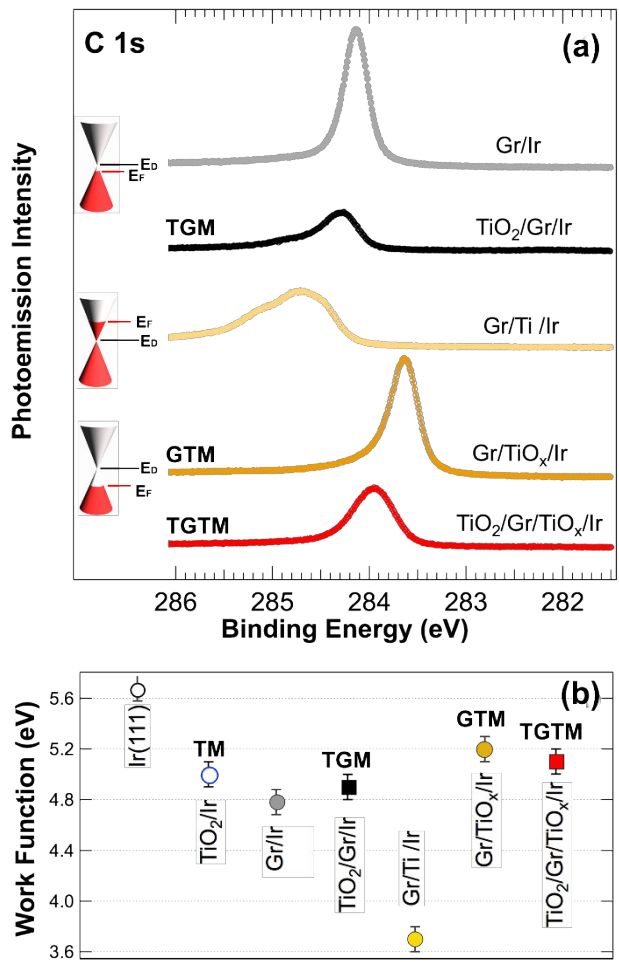

**Figure 4.** Charge redistribution in the different nano-architectures. a) C *1s* core level photoemission spectra measured on selected systems (hν = 400 eV). The approximate position of the Fermi level $E_F$ with respect to the Dirac cone $E_D$ is also reported; b) Comparison of the work function measured on the different nano-architectures.

These differences in the Gr substrate doping are expected to affect the $TiO_2$ NPs by modifying their electronic level occupation, and thereby their WF. Therefore, we have assessed how the presence of Gr and its substrate affect the WF of the different nanoarchitectures. Figure 4b displays the WF of the three substrates, before (circles) and after (squares) the growth of $TiO_2$ NPs. The WF of titania is close to the value of 5.1 eV, reported for bulk anatase, in all three systems. However, there are differences of a few hundred meV between them. Firstly, the WF of the Ir- (grey circle) and $TiO_x$- supported (orange circle) Gr clearly reflects the shift observed in the C *1s* core level, as already reported [48]: the WF of the former is indeed lower than that of the latter by about 400 meV. It is interesting to highlight that the WF of the two Gr substrates (black and red squares) also affects that of the NPs supported on them. In particular, it is evident that their WF lies in between that of bulk $TiO_2$ and that of the respective substrate, indicating a charge transfer to or from the NPs depending on the degree of Gr doping.

Following this characterization, we qualitatively tested the photocatalytic activity of the different interfaces for a sample reaction, i.e. the hydrogen evolution half-reaction. To this aim, the samples were immersed in a 1:1 water/methanol solution and exposed to the radiation of a solar simulator. Methanol is conventionally used as sacrificial reagent, that can interact with the photogenerated holes and be oxidized more easily with respect to water, avoiding the partial back recombination of $H_2$ and $O_2$ formed during pure water splitting. In this respect, these measurements are meant to show qualitative trends between our systems, while an evaluation of the actual performance of these catalysts is outside the scope of our experiment. Nevertheless, for the sake of completeness, we provide reference measurements taken in the same conditions on a conventional Gr/Titania catalyst in the Supporting Information. The activity of the nano-architectures was evaluated by measuring the amount of $H_2$ gas produced in 20 hours. The results of the measurements of the TGM and TGTM systems are reported in Table 1. It is interesting to note that a significant effect is observed when comparing the activity of titania on metal- (TGM) and oxide-supported Gr (TGTM). The latter system has an activity that is much larger than the former, thus clearly showing that the presence of a two-dimensional $TiO_x$ oxide in between the substrate metal and monolayer Gr plays a vital role in determining the photocatalytic efficiency of the supported catalyst.

**Table 1.** Photocatalytic measurements. Hydrogen production, expressed in [mol $H_2$/mol $TiO_2$], from a water/methanol 1:1 solution in 20 hours under simulated solar illumination by the TGM and TGTM nano-architectures, normalized by the quantity of titania.

| System | TGM | TGTM |
|---|---|---|
| $H_2$ production [mol $H_2$/mol $TiO_2$] | 0.103 ± 0.008 | 0.835 ± 0.005 |

To explain the origin of these large differences among the various nano-architectures and to understand the fundamental role of the $TiO_x$ surface oxide at the interface, we performed DFT calculations for the TGM and TGTM architectures. The systems were described with a slab of 4 layers of Ir in a 10 × 10 hexagonal supercell. The bottom two layers of Ir were kept frozen at their bulk geometry, with a lattice parameter of 2.74 Å. A $TiO_x$ layer was initially placed on top of Ir for the case of the TGTM system, using the geometry experimentally observed for a monolayer of oxidized Ti on the Pt(111) surface having this stoichiometry. In both TGM and TGTM systems, (11 × 11) unit cells of graphene were superimposed. This definition of the supercell was able to describe the (2 × 2) periodicity of the $TiO_x$ layer below graphene. This (11 × 11) periodicity for graphene, which corresponds to a (10 × 10) Ir supercell, does not exactly reproduce the experimentally observed one, but it results in a graphene lattice strain that is less than 1%. Eventually, titania layers in the anatase configuration with (001) and (101) surface terminations were placed on top of Gr, with a thickness of 7.94 and 9.70 Å, respectively, i.e. very similar to the one measured by STM. The total number of atoms considered for the TGTM system was 1397 (400 Ir, 260 Ti, 242 C and 495 O).

Interestingly, the (101) surface of anatase (see Figure 5a and b), which is known to be the most efficient in the reduction reactions [54] because of the electron trapping at the aqueous-surface interface [55], resulted to be energetically more stable by 35 eV, in agreement with the Wulff construction [56]. The (101) titania anatase surface exhibits the typical sawtooth-like corrugation with 6- and 5-fold coordinated Ti, and 3- and 2-fold coordinated O atoms [53]. On the contrary, the finite thickness of the (100) slab results in a large deformation of the anatase structure (see Supporting Information), indicating that the tendency of this surface to reconstruct [57] is also present in the case of a thin film supported by Gr. The introduction of a $TiO_x$ surface-oxide in between Gr and the metal substrate almost entirely removes the 0.32 Å corrugation of the carbon layer which is present in the TGM structure. For the two layered structures, the titania-Gr distance is not appreciably different (see Figure 5a and b).

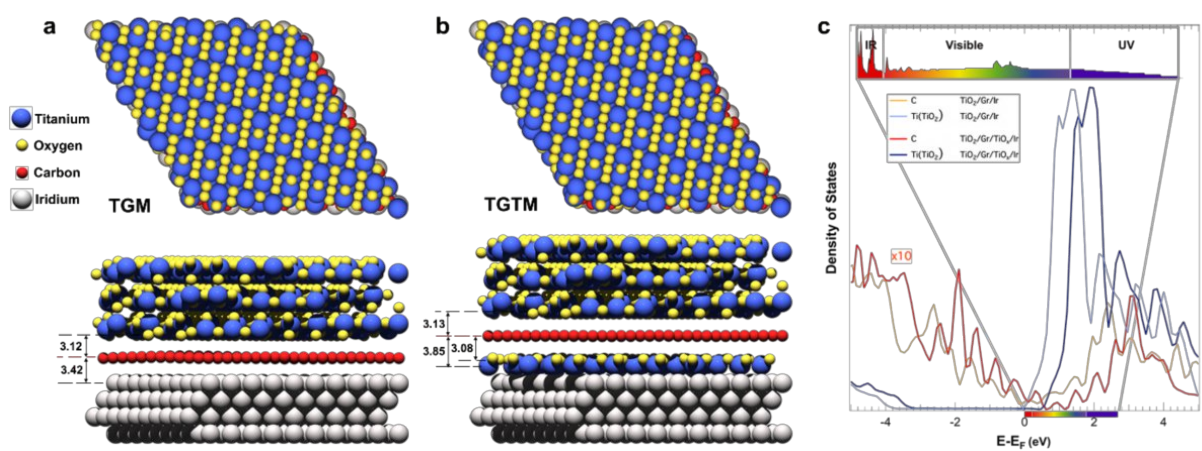

**Figure 5.** Geometric and electronic structure of the TGM and TGTM nanoarchitectures from DFT calculations. Top and side view of the a) relaxed TGM and b) TGTM supercells; c), calculated partial density of states of TGM (yellow and light blue curves are multiplied by a factor 10) and TGTM (red and blue curves) with (101) surface termination.

To understand the role of the interfacial oxide, we computed the density of states (DOS) projected on the carbon atoms and on the titanium atoms of the $TiO_2$-anatase layer. Besides the band gap of 3.2 eV, which is in good agreement with the optical gap of anatase measured at 4 K [58], Figure 5c clearly shows that both the valence and conduction bands of $TiO_2$ are shifted by a similar amount (about 0.5 eV) to higher energies for the TGTM architecture. Interestingly, the same overall upward shift of 0.5 eV can be appreciated also in the π and π* bands of Gr (red curve), as a main perturbation caused by the addition of the $TiO_x$ interfacial oxide. This clearly reflects the experimentally observed effect of the Gr doping and work function of Gr on the work function of the NPs. In particular, because of the Gr p-doping, the bottom of the titania conduction band moves away from the Fermi level, with the important effect of a region just above this level characterized by a zero density of empty states, on the contrary to what is found for the TGM structure where the lower edge of the Ti $3d^*$ band lies very close to the Fermi level and the $TiO_2$ is almost degenerate. In addition, the higher separation of these states from the Fermi level in the TGTM system has a dramatic effect on the de-excitation channels available to the electrons excited into them. In particular, one of the most probable decay channels, i.e. a phonon-assisted process, requires a much larger number of electron-phonon scattering events the further the states are from the Fermi level. Therefore, the higher energy of the Ti $3d^*$ conduction band states considerably reduces the e-h pairs recombination in Gr in TGTM when compared with TGM. Further contributions that might explain the enhanced reactivity of the novel nano-architecture can be found in the higher Gr DOS at the Fermi level and in the reduction of the Gr DOS above the Fermi energy. The former increases the number of excited electrons, the latter reduces the possibility of electron excited in the conduction band of Gr to recombine with holes via phonon-assisted processes. All these contributions affecting the excitation and de-excitation channels of e-h pairs allow this p-doped nanostructure to show a dramatically increased performance despite its chemical state and geometrical structure not showing any significant difference.

## 3. Conclusion

In summary, we have shown that a layered nanoarchitecture formed by titania NPs, high-quality single-layer graphene and two-dimensional Ti oxide film supported by a metal substrate shows an enhanced photocatalytic activity, which is controlled by the degree of Gr doping and by the shifts of Ti $d$ band.

The change in the electronic structure of the layered material strongly reduces the probability of recombination of electrons and holes which are produced in the photoexcitation process. Our results confirm that the use of 2D materials, already successfully applied in the case of solar-driven conversion of $CO_x$ [59], and in particular of transfer-free epitaxially grown materials and their heterostructures could be applied for the design of novel energy related materials with greatly improved functionalities.

## 4. Methods

4.1 *Experimental methods*

The Ir(111) surface was cleaned by repeated cycles of $Ar^+$ sputtering and flash annealing to 1400 K, followed by annealing in $O_2$ and in $H_2$ gas [60]. The single-layer Gr growth on Ir(111) has been performed by repeated annealing cycles of the sample to 1420 K while exposing it to up to $3\times10^{-7}$ mbar partial pressure of ethylene. We checked the quality of Gr by low-energy electron diffraction and scanning tunneling microscopy, to evaluate its long-range order and density of point-like defects, and high-resolution photoelectron spectroscopy. The titanium oxide intercalation below graphene was achieved with a two-step procedure. Firstly, we exposed the graphene-covered surface to Ti sublimated from a high-purity filament, while keeping the sample at 670 K. The dose of Ti deposited was monitored by using a quartz microbalance. Afterwards, we exposed the sample to $5\times10^{-3}$ mbar partial pressure of oxygen for one hour, while keeping it at 570 K. We employed a two-step procedure for the NPs growth on graphene, keeping the sample at room temperature and exposing it to Ti atoms first and then to a partial pressure of $1\times10^{-6}$ mbar of oxygen for ten minutes. High-resolution (HR) X-ray photoelectron spectroscopy (XPS) measurements were performed in-situ at the SuperESCA beamline at Elettra. The experimental chamber is equipped with a Phoibos hemispherical electron energy analyzer, provided with a delay line detector. The overall energy resolution was always better than 100 meV for the photon energies and parameters employed. The XPS spectra were acquired by tuning the photon energy in order to have a photoelectron kinetic energy of about 100 eV, to enhance surface sensitivity. For each spectrum, the photoemission intensity was normalized to the photon flux and the binding energy (BE) scale was aligned to the

Fermi level of the iridium substrate. For the fitting procedure of the core levels, a Doniach-Šunjić line profile has been used for each spectral component, convoluted with a Gaussian distribution to account for the experimental, phonon and inhomogeneous broadening [61]. The background was modeled with a polynomial of first order for all spectra, except for those of the Ti *2p*, where a second order polynomial was used. Resonant Photoelectron Spectroscopy was performed by scanning across the Ti $L_3$-edge – between 452 and 462 eV photon energy – while measuring the valence band spectral region. The photon energy was varied by 0.1 eV at each step and the photon energy calibration was performed by using the second order diffraction of the monochromator. The on-resonance photon energy was defined in coincidence to the maximum photoemission intensity in the valence band region at 458.8 eV. For the off-resonance a photon energy of 451 eV was selected. The surface work function (WF) was evaluated by measuring the kinetic energy onset of the secondary electrons signal, using a (previously calibrated) photon energy of 140 eV. In order to be able to carry out this measurement also for samples having a lower WF than the analyzer, a bias of −10 V was applied to the sample in the measurements.

Near edge X-ray absorption fine structure (NEXAFS) measurements have been performed in Auger yield mode at the SuperESCA beamline, normalizing the intensity to the photon flux for each photon energy. The photon flux has been measured via the total drain current from a gold mesh intercepting the photon beam.

Scanning tunneling microscopy (STM) measurement have been performed ex-situ at the CoSMoS experimental station of the SuperESCA beamline using a SPECS STM 150 Aarhus equipment. The images reported in this article have been acquired using a tungsten tip, at constant current (0.47 nA) with 2.1 V bias and at room temperature. The calibration of the perpendicular coordinate has been done by measuring the already known height of the step between iridium surface and Gr. For the image analysis, we used the free open-source software Gwyddion [62]. The statistical analysis on the cluster size has been performed using the specific built-in functions.

The photocatalytic activity was evaluated under simulated sunlight irradiation using a solar simulator (LOT-Oriel) equipped with a 150 W Xe lamp and an atmospheric edge filter to cut-off UV photons below 300 nm. The beam was focused on the sample and the resulting light intensity was 25 mW cm$^{-2}$ (250 – 400 nm, UV-A) and 180 mW cm$^{-2}$ (400 – 1000 nm, Vis-NIR). The incident illumination power is close to 2 Suns and is representative of a simple but effective solar concentrator. The photocatalytic hydrogen evolution reaction (HER) experiments were performed with head spaced vials (total volume 20 mL) filled with 12.5 mL of a water/methanol 1:1 v/v solution. The vial was then sealed using appropriate rubber septa and the air was removed by bubbling Ar for 30 minutes. After this equilibration period, the sample was irradiated for 20 hours at room temperature. The analysis of the reaction products was performed injecting 50 µL of the gas phase into a gas chromatograph (Agilent 7890), after adding 250 µL as internal standard. The Thermal Conductivity Detector (TCD) was used for the quantification of $H_2$, using a MolSIEVE 5A column with Ar as carrier.

*4.2 Theoretical methods*

The calculations have been performed using density functional theory (DFT) as implemented in the VASP code [63]. The atomic structure of the studied systems was fully relaxed using the rev-vdw-DF2 functional until the largest residual force was less than 0.015 eV/Å [64]. We employed the projector augmented method (PAW) [65], using PBE potentials [66], with 9, 4, 4 and 6 electrons in valence for Ir, Ti, C and O, respectively. The plane wave cutoff was set to 400 eV, and the relaxations were performed by sampling the Brillouin zone using the Γ point only. Although the geometry of the system is accurately described by the rev-vdw-DF2 functional, its electronic structure requires the use of a hybrid functional, as pointed out earlier [67]. To obtain the partial density of states (PDOS) we have therefore performed single point DFT calculations with the HSE06 functional [68], using geometries obtained with the rev-vdw-DF2 functional. Because of the size of the system, each HSE06 calculation required several weeks of running on 768 cores of a Cray-XC30 supercomputer. By contrast, similar calculations with the rev-vdw-DF2 functional only took a few minutes. With our supercell, the PDOS of Gr is poorly described by the Γ point only, requiring at least a 3 × 3 Monkhorst-Pack k-point grid [69], but the PDOS of Ti is already very accurate with just the Γ point.

## Acknowledgments

We are grateful, for computational resources, to ARCHER UK National Supercomputing Service, United Kingdom (NE/M000990/1 and NE/R000425/1), the University College London (UCL) Research Computing, the MMM hub (EP/P020194/1) and Oak Ridge Leadership Computing Facility (DE-AC05-00OR22725). D. De Angelis and F. Presel contributed equally to this work.

## Conflict of Interest

The authors declare no conflict of interest.

## Appendix A. Supplementary Data

Supplementary data to this article can be found online at

Supplementary data for

# Interfacial two-dimensional oxide enhances photocatalytic activity of graphene/titania via electronic structure modification

Dario De Angelis[a,†], Francesco Presel[a,†,h], Naila Jabeen[a,b,c], Luca Bignardi[d,i], Daniel Lizzit[d], Paolo Lacovig[d], Silvano Lizzit[d], Tiziano Montini[e], Paolo Fornasiero[e], Dario Alfè[f], Alessandro Baraldi[a,d,g,*]

[a] Department of Physics, University of Trieste, via Valerio 2, 34127 Trieste, Italy.

[b] Nanosciences & Catalysis Division, National Centre for Physics, Islamabad 44000, Pakistan.

[c] International Centre for Theoretical Physics, Strada Costiera 11, 34151 Trieste, Italy.

[d] Elettra-Sincrotrone Trieste, S. S. 14, km 163.5 in AREA Science Park, 34149 Trieste, Italy.

[e] Department of Chemistry and Pharmaceutics, University of Trieste, and INSTM, via L. Giorgieri 1, 34127 Trieste, Italy.

[f] Department of Earth Sciences, Department of Physics and Astronomy, Thomas Young Centre@UCL, London Centre for Nanotechnology, University College London, Gower Street, WC1E 6BT London, United Kingdom.

[g] IOM-CNR, Laboratorio TASC, S. S. 14 km 163.5 in AREA Science Park, 34149 Trieste, Italy.

[†] These authors contributed equally.

* Corresponding author. E-mail: alessandro.baraldi@elettra.eu (Alessandro Baraldi)

[h] Present address: Department of Physics, Technical University of Denmark, 2800 Kgs. Lyngby, Denmark.

[i] Present address: Department of Physics, University of Trieste, via Valerio 2, 34127 Trieste, Italy.

**S1. Valence Band and Resonant Photoemission spectra**

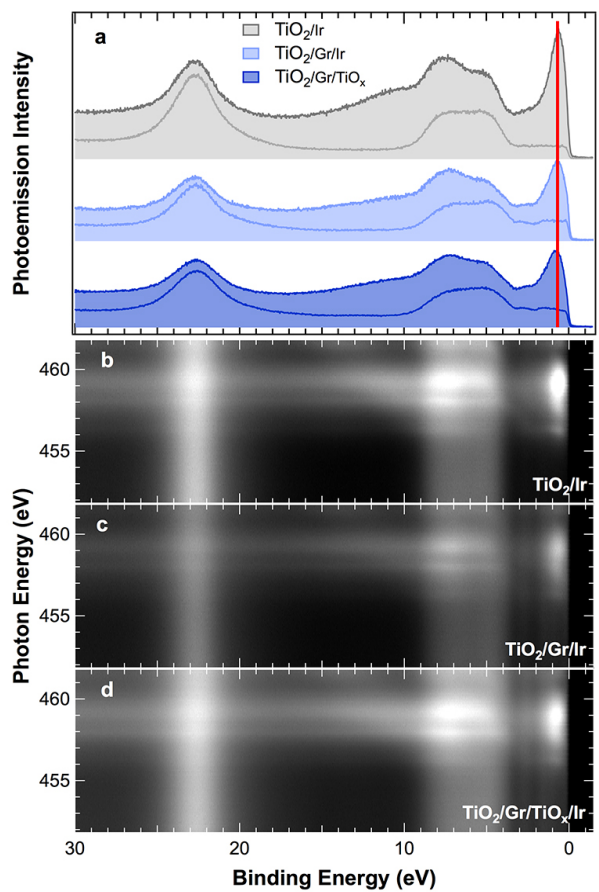

**Figure S1:** Valence Band and Resonant Photoemission spectra on TM, TGM and TGTM systems. a) Valence band spectra acquired on- and off- resonance (458.8 eV and 451 eV photon energy respectively) at the Ti $L_3$-edge. Resonant Photoemission Spectroscopy data for b) TM, c) TGM, and d) TGTM, measured across Ti $L_3$-edge. The photoemission intensity is plotted in grayscale, dark representing lower photoemission intensity.

In the RESPES measurements, we observed main spectral features, which show a resonant enhancement in their photoemission intensity for photon energies between 458 and 460 eV, attributed to hybridized O 2$p$-Ti 3$d$ states and to the O 2$s$ level, and a narrow component close to the Fermi level (1.1 eV). Such feature is widely recognized as a fingerprint of the defect states introduced into the band gap by the self-doping of titania by Ti atoms in a lower oxidation state due to an adjacent oxygen vacancy [1].

## S2. Characterization of the reference powder composite $TiO_2$/graphene

As a reference, the photocatalytic activity was also tested in the same conditions for a conventional Ti/Gr powder.

A $TiO_2$/graphene composite in powdered form was prepared by sol-gel technique starting from a commercial graphene material (Thomas-Swan, Premium Grade Graphene Powder). 3.10 g of Ti(OBu)$_4$ polymer were dissolved in 8.0 mL of ethanol. 84 mg of graphene were dispersed in this solution by ultrasound sonication for 2 hours. Hydrolysis of the titanium precursor was realized by adding a solution containing 4.1 mL of $HNO_3$ 65%, 2.2 mL of $H_2O$ and 4.4 mL of ethanol. The suspension was maintained under vigorous stirring until a gel is formed. The gel was aged at room temperature for 24 h and then dried at 353 K overnight. Finally, all the organic and inorganic residues were removed by calcination in air at 523 K for 6 hours. The obtained material was characterized by Thermogravimetric analysis, Raman spectroscopy and XPS. Thermogravimetric analysis (TGA) was performed using a TGA Q500 (TA Instruments) under air flow (100~mL/min), by equilibrating at 373 K for 20 minutes and then ramping the temperature at 10 K~min$^{-1}$ up to 1173 K (Figure S3). Raman spectra are recorded with an inVia Renishaw microspectrometer equipped with Nd:YAG laser at 532 nm. To avoid sample damage or laser-induced heating/crystallization of the materials, the incident power was kept at 1% (full power of the laser is 100 mW). Powders are dispersed in EtOH, drop-cast onto a quartz slide and the solvent evaporated. The chemical state of the elements in the composite materials was determined by XPS analysis performed in the Surface Science Laboratory of Elettra-Sincrotrone Trieste. The spectra were acquired using a VG Escalab electron energy analyser instrument and a Mg K$\alpha$ x-ray source at 1253.6 eV, with an overall resolution of 1.2 eV.

The weight of the material slowly decreases during heating in flowing air up to 500°C, with a loss around 2.5 wt% corresponding to the desorption of adsorbed water and $CO_2$ (Figure S2). At higher temperature (500 – 900 °C), a sharper weight loss resulting from graphene burning is observed. From this data, it is possible to estimate a graphene content of 4.9 wt% in the powdered composite photocatalyst.

The Raman spectrum of the TiO$_2$/graphene sample (Figure S3) shows the typical features of TiO$_2$ anatase (bands at 149, 402, 518 and 641 cm$^{-1}$) and of graphene (D band at 1355 cm$^{-1}$, G band at 1586 cm$^{-1}$ and 2D band at 2721 cm$^{-1}$). The spectrum shows the presence of a high-intensity D-band component, which is related to the disorder and defects in graphene.

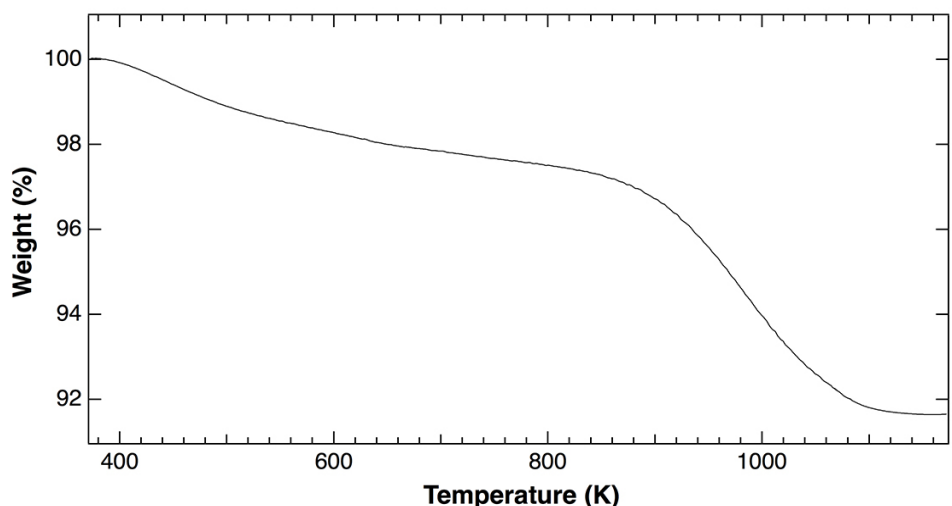

**Figure S2.** Thermogravimetric analysis (TGA) for the TiO$_2$/graphene powdered composite.

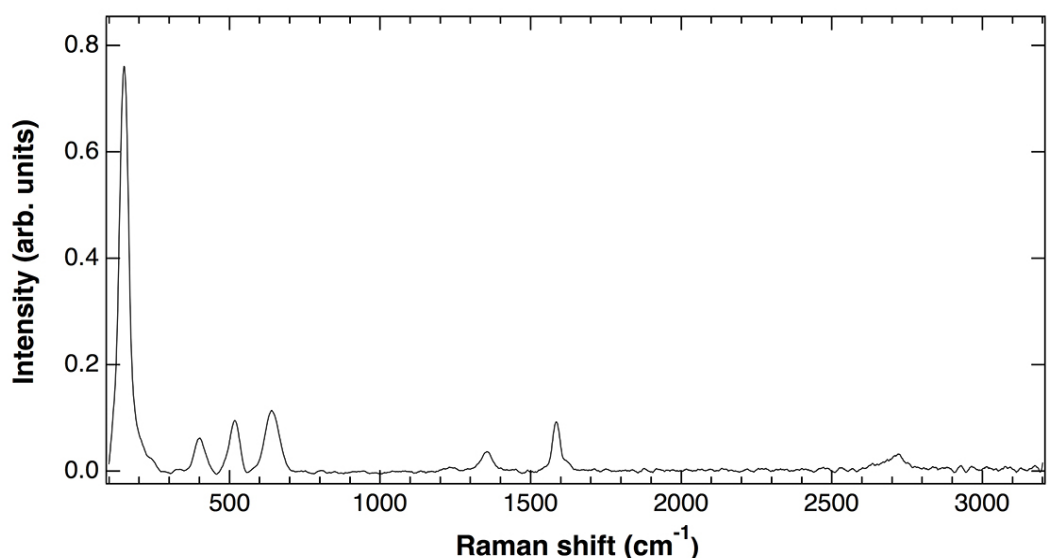

**Figure S3.** Raman spectrum of the TiO$_2$/graphene powdered composite.

The survey spectrum and the spectral region of the C 1s, O 1s and Ti 2p levels are presented in Figure S4. The C 1s signal presents various components, with the most intense ones attributed to sp$^2$ C (284.5 eV) and C-O groups (286 eV) and a small signal deriving from carboxyl groups (290 eV) and the Ti-O-C interface [2]. This indicates that the RGO is only partially reduced, with a high density of functional groups. The traces of N detected in the survey can also be attributed in impurities in the RGO. The O 1s signal shows the typical pattern dominated by the signal from surface OH groups (531.5 eV), as expected given the low calcination temperature of the material. Finally, the Ti 2p spectrum presents one single component at about 459 eV, corresponding to Ti$^{4+}$.

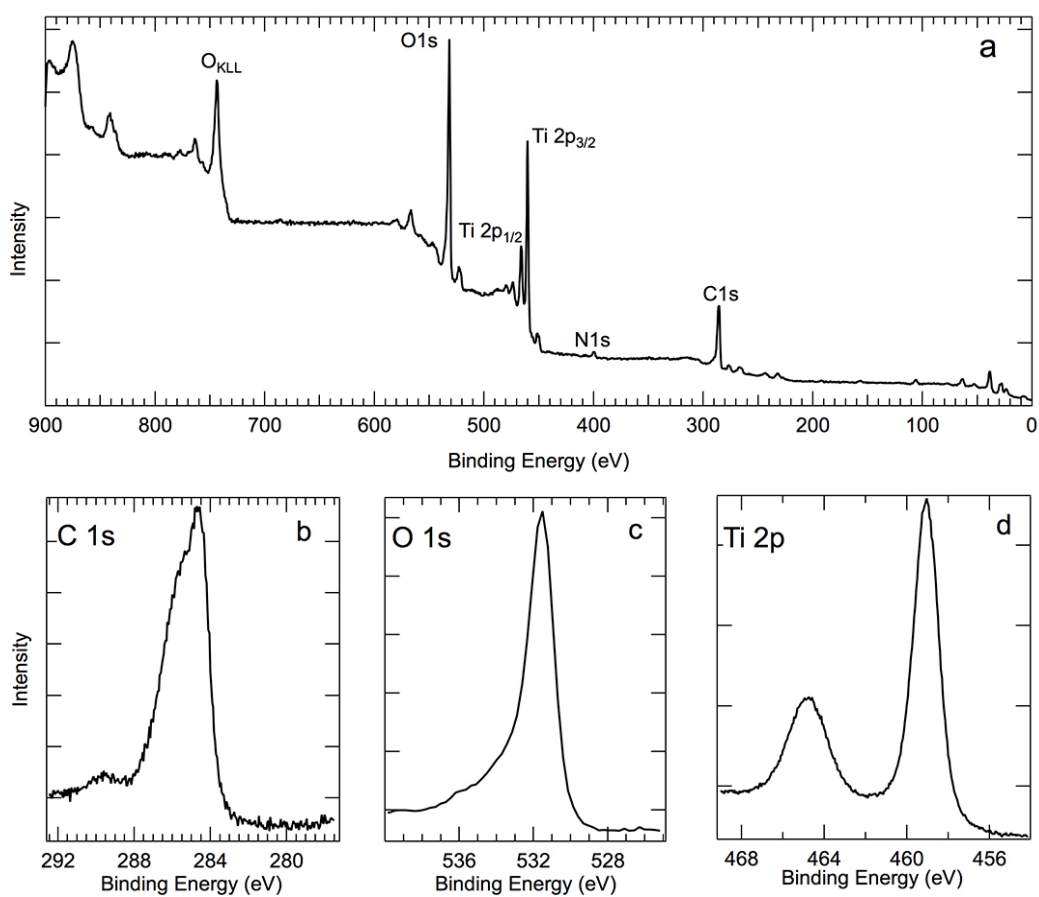

**Figure S4.** X-ray Photoelectron Spectroscopy measurements of the TiO$_2$/graphene powdered composite. a) Survey spectrum. b) C 1s spectrum. c) O 1s spectrum. d) Ti 2p spectrum.

The following table shows the results of the photocatalitic activity measurements for the TiO$_2$/Ir interface (TM), in absence of the Gr support, and that of the reference powder, compared to the Gr-supported systems.

| System | TM | TGM | TGTM | powder |
|---|---|---|---|---|
| H2 prod. [mol H2 /mol TiO2 ] | 0.035 ± 0.009 | 0.103 ± 0.008 | 0.835 ± 0.005 | (1.28 ± 0.02) × 10−4 |

By comparing the first two systems, it is confirmed that the use of Gr as support has a significant effect on the activity of titania, which shows an almost three-fold increase when the titania is supported on Gr/Ir (TGM) rather than on the metal substrate only. It must be stressed that the capability to act as electron acceptor depends on its functionalization/ presence of defects. The use of high-quality single-layer Gr is therefore of paramount importance for enhancing the performance.

Notably, all the nanostructured materials present significantly higher activity per gram of TiO$_2$ with respect to the conventional powder reference sample, and in particular the TGTM interface demonstrated an activity for H$_2$ evolution which is orders of magnitude higher than the powder TiO$_2$/Gr. In fact, since photocatalysis depends on photon flux that reaches the semiconductor and on the molecules adsorbing on the surface, both the electronic properties of the investigated material and its surface area are of great importance. This result highlights how the fine control of the interface is crucial to obtain straightforward performances in the H$_2$ evolution semireaction.

**S3. DFT calculations: anatase titania (100) surface termination**

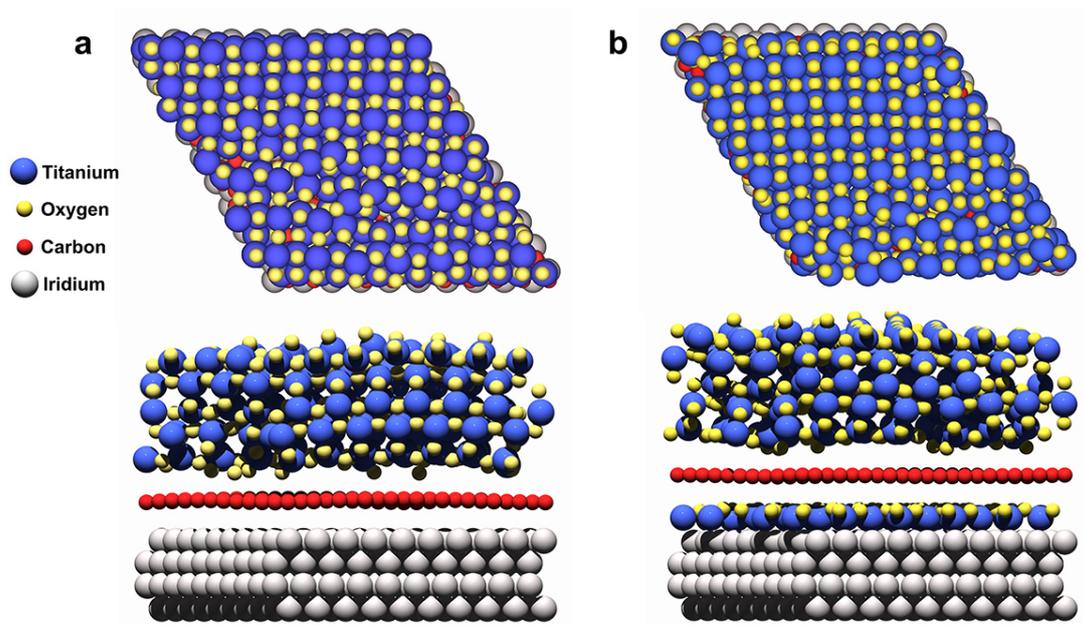

**Figure S5.** Structure of the TGM and TGTM nanoarchitectures corresponding to the (100) anatase TiO$_2$ surface termination from DFT calculations. a) Top and side view of the relaxed TGM supercell. b) Top and side view of the relaxed TGTM supercell.